\documentclass[%
reprint,
12pt,
amsmath,amssymb,
aps,
onecolumn,
showkeys,
pra,
]{revtex4-2}

\usepackage{color}
\usepackage[hoptionsi]{overpic}

\usepackage{graphicx}      
\usepackage{dcolumn}       
\usepackage{bm}            
\usepackage{amsmath}
\usepackage{amssymb}
\usepackage{cancel}
\usepackage{esint}
\usepackage{fixmath}
\usepackage{algorithm}
\usepackage{algorithmic}
\usepackage{theorem}
\usepackage{comment}

\usepackage{geometry}
\usepackage{setspace}
\usepackage{float}

\usepackage{booktabs}
\usepackage{svg}
\usepackage{xcolor}
\usepackage[export]{adjustbox}
\usepackage{pifont}

\usepackage{tikz}
\usetikzlibrary{arrows.meta}
\definecolor{shockblue}{RGB}{0,142,184}
\definecolor{bodyfill}{RGB}{243,245,246}
\definecolor{bodyline}{RGB}{30,30,30}

\usepackage{caption}
\usepackage{subcaption}
\usepackage{ragged2e}
\DeclareCaptionJustification{justified}{\justifying}

\usepackage{hyperref}

\newcommand {\be}{\begin{equation}}
\newcommand {\ee}{\end{equation}}
\newcommand {\bse}{\begin{subequations}}
\newcommand {\ese}{\end{subequations}}
\newcommand {\bs}[1]{\boldsymbol{#1}}
\newcommand {\bq}{\bs{q}}
\newcommand {\bQ}{\bs{Q}}

\newcommand {\bL}{\bs{\Lambda}}
\newcommand {\bPi}{\boldsymbol{\Pi}}
\newcommand {\Dxi}{\mathcal{D}_\xi}

\newcommand {\td}{\text{d}}
\newcommand {\tT}{\text{T}}
\newcommand {\f}{\frac}
\newcommand {\p}{\partial}

\begin{document}

\title{Receptivity of the flow on the stagnation streamline of a blunt body in supersonic flow}

\author{Iliya Milman}
\author{Michael Karp\textsuperscript{\dag}}

\affiliation{The Stephen B. Klein Faculty of Aerospace Engineering, Technion Israel Institute of Technology, Technion City, Haifa, Israel}

\affiliation{\vspace{0.2em}\rm{\textsuperscript{\dag}Corresponding author: \href{mailto:mkarp@technion.ac.il}{mkarp@technion.ac.il}}}

\begin{abstract}
The receptivity of the inviscid flow on the stagnation streamline of a blunt body in supersonic flow is investigated theoretically for incoming freestream disturbances. 
The wave transmission and coupling are quantified by solving the linearized shock-fitting problem with a spectral method, whereas the steady base flow is obtained using a nonlinear shock-fitting spectral solver. Revisiting previous theoretical work, we identify and correct an error in a key coefficient in the analysis by Morkovin ({\it{J. Appl. Mech.}}, {\bf{27}}, 1960), overturning the prior conclusion of body-induced damping and revealing amplification instead. The post-shock entropy disturbances display singular behavior near the stagnation point, which is treated analytically. Acoustic disturbances dominate pressure and velocity responses, while density is affected by both acoustic and entropy modes. The base flow pressure gradient introduces weak coupling between the acoustic and entropic components of the response. The actual stagnation-line base flow amplifies all disturbances more than the simplified model of uniform post-shock flow; as well as a shock without a body, and the differences are quantified for a range of Mach numbers. The responses to entropy, fast acoustic, and slow acoustic waves are compared as functions of the freestream Mach number.
\end{abstract}


\maketitle

\section{Introduction}

Understanding the interaction between freestream disturbances and shock waves is central to predicting unsteady behavior in hypersonic blunt body flows. \citet{Ribner54a,Ribner54c} conducted the foundational work on how disturbances interact with shock waves. Through a linearized analysis of normal shock jump conditions, he demonstrated that each type of infinitesimal upstream disturbance - acoustic, entropy, or vorticity - generates all three disturbance modes upon interaction with the shock. This coupling underscores the complex nature of shock-disturbance interactions. \citet{Moore54} extended this theoretical framework to oblique shocks, while \citet{McKenzieWestphal} derived analytical expressions for the amplitudes of transmitted and refracted waves, showing how the transmission, reflection, and mode generation coefficients depend both on the Mach number and the angle of incidence.

\citet{morkovin1960note} was among the first to include reflections from a body downstream of the shock. He examined the flow on the stagnation line to assess how upstream disturbances - particularly entropy fluctuations - generate downstream perturbations. Morkovin discussed the possibility of a resonance between the transmitted acoustic and entropy waves at the shock, and the reflected acoustic wave. While reproducing his results, we discovered an error in his paper: although the coefficient $\Pi_{31}$ is correctly written in its explicit form, its sign is consistently incorrect in all the plots, i.e. a factor of $-1$ is missing when $\Pi_{31}$ is plotted. This sign error leads to the misleading conclusion that the body dampens the disturbances, which is puzzling, since reflections are expected to amplify the disturbances. In addition, his analysis assumes that the post-shock flow is uniform, which is unrealistic, as the actual flow along the stagnation line decelerates to a stop. Our analysis reproduces Morkovin's results using the correct sign of the coefficient $\Pi_{31}$ and quantifies the differences in disturbance amplification between the actual flow and the idealized uniform flow. Additional discussion of the analysis by \citet{morkovin1960note}, including a correction of its figures, is given in Appendix~\ref{AppA}.

Building on these foundations, recent studies have explored amplification mechanisms within the region between the bow shock and the body. \citet{Chaudhry2017} employed direct numerical simulations alongside an analytical model to characterize the transfer function between freestream disturbances and pitot pressure spectra in hypersonic flows. Their results revealed that a resonance between the shock and probe geometry can significantly amplify pressure fluctuations. \citet{SchildenJFM17,SchildenJFM19} showed that stagnation point probe signals in supersonic flow are strongly affected by the type, frequency, and wave inclination of the freestream disturbance, underscoring the need for careful probe calibration. While the present study follows previous works and focuses on the response of the shock to freestream perturbations, recent findings by \citet{sawant22}, considering the internal structure of the shock using kinetic approaches, point to the existence of shock internal frequencies which may trigger different receptivity mechanisms. More recently, \citet{Xiong2023} used a one-dimensional model to demonstrate that post-shock disturbance resonance can lead to substantial amplification under certain flow conditions. Furthermore, simulations of unsteady hypersonic flow over blunt bodies using shock-fitting techniques that accurately capture the stagnation line, such as those of \citet{brooks2004} and \citet{najafi2014shockfit}, have neglected the generated entropy disturbances along the stagnation line, without offering a detailed justification for this simplification. 

The objective of the present work is to examine the behavior of infinitesimally small disturbances along the stagnation line of a blunt body. Our theoretical analysis combines the shock-fitting approach with linear stability theory (LST) across a broad parameter space, enabling a more complete understanding of shock-generated disturbances and their downstream effects. The difficulties of capturing the generated entropy disturbances are discussed and quantified. In addition, the role of the base flow pressure gradient in coupling between the post-shock acoustic and entropy disturbances is analyzed.

\section{Mathematical model}\label{sec:math}

\subsection{Base Flow}\label{2.1}

The base flow is a steady solution of the two-dimensional Euler equations that describe the isentropic flow of an inviscid, calorically perfect gas in the shock layer (the region between the bow shock and the body). The base flow is computed using a shock-fitting algorithm with a high-accuracy spectral discretization in space, which calculates the shock shape as part of the solution, as described in detail in \citet{MilmanKarp}, as well as previous works by \citet{kopriva1999} and \citet{brooks2004}. The unsteady Euler equations in the shock layer are marched in time until convergence to steady state. The Rankine-Hugoniot relations provide the jump conditions at the moving shock, while its acceleration is determined by a compatibility equation that accounts for waves propagating from the shock layer back to the shock. At the body surface, the no-penetration boundary condition is prescribed. Further details, including the governing equations, are provided in \citet{MilmanKarp}. The flow variables are nondimensionalized as follows,

\be\label{nondim}
t = \f{{t}^*\sqrt{p_\infty/\rho_\infty}}{L},\,
x = \f{{x}^*}{L} ,\,
y = \f{{y}^*}{L} ,\,
u = \f{{u}^*}{\sqrt{p_\infty/\rho_\infty}} ,\,
v = \f{{v}^*}{\sqrt{p_\infty/\rho_\infty}} ,\,
p = \f{{p}^*}{p_\infty} ,\,
\rho = \f{{\rho}^*}{\rho_\infty},
\ee
where $u$ and $v$ are the horizontal ($x$) and vertical ($y$) velocity components, respectively, $p$ is the pressure and $\rho$ is the density. The asterisk indicates dimensional quantities, $M_\infty$ is the freestream Mach number, $p_\infty$ is the freestream pressure, $\rho_\infty$ is the freestream density and $L$ is a reference length, which for a blunt body is usually chosen as the leading edge radius. In addition, the entropy is defined as

\be
s = \f{{s}^*}{c_v} = \ln\f{p}{\rho^\gamma},
\ee
where $c_v$ is the specific heat capacity at constant volume, and $\gamma$ is the adiabatic index, with $\gamma=1.4$ throughout this paper.

\subsection{Equations on the stagnation streamline}

Along the stagnation streamline, the vertical velocity component vanishes, i.e. $v = 0$, due to symmetry (however, $\p v/\p y$ is retained for now). In this case, the continuity equation is replaced by the entropy convection equation. Under these assumptions, the governing equations simplify to

\bse
\be
\f{\p u}{\p t} + u \f{\p u}{\p x} + \f{1}{\rho}\f{\p p}{\p x} =0,
\ee
\be
\f{\p p}{\p t} + u \f{\p p}{\p x} + \gamma p \left(\f{\p u}{\p x} +\f{\p v}{\p y}\right) = 0,
\ee
\be
\f{\p s}{\p t} + u \f{\p s}{\p x} = 0.
\ee
\ese
The shock-fitting methodology is utilized to map the physical domain, $(x, t)$, to the computational domain, $(\xi, \tau)$, where the Chebyshev-Gauss-Lobatto distribution is used. The transformation is given by

\be\label{transform0}
\xi = 1 - 2 \f{x - x_b}{x_s(t) - x_b},\quad  \tau = t,
\ee
where $x_b$ represents the stationary body coordinate, and $x_s(t)$ denotes the time-dependent shock coordinate. Thus, the stagnation streamline, $x\in[x_s,x_b]$, is mapped to $\xi\in[-1,1]$. Applying these transformations, the equations in the computational domain are given by

\bse\label{eq:nonlinear}
\be
\f{\p u}{\p \tau} + (\xi_t + u \xi_x) \f{\p u}{\p \xi} + \f{1}{\rho} \xi_x \f{\p p}{\p \xi} =0,
\ee
\be
\f{\p p}{\p \tau} + (\xi_t + u \xi_x) \f{\p p}{\p \xi} + \gamma p \left(\xi_x\f{\p u}{\p \xi} +\f{\p v}{\p y}\right) = 0,
\ee
\be
\f{\p s}{\p \tau} + (\xi_t + u \xi_x) \f{\p s}{\p \xi} = 0,
\ee
\ese
with the following transformation metrics, 

\be
\xi_t = \f{\xi - 1 }{x_b-x_s} \f{\td x_s}{\td\tau},\quad \xi_x = \f{2}{x_b-x_s}.
\ee


\subsection{Linear stability analysis}

The above equations are linearized by decomposing the variables $\bq=(u,p,s,x_s)^\tT$, namely velocity, pressure, entropy, and shock location, respectively, into a steady base flow $\bQ=(U,P,S,X_s)^\tT$ and a disturbance $\bq'$,

\be\label{decomp}
\bq = \bQ(\xi) + \varepsilon \bq'(\xi,\tau), \quad \varepsilon\ll 1.
\ee
The base flow is a steady solution of \eqref{eq:nonlinear}, i.e.,

\be\label{baseflow_eqs}
R U \f{\td U}{\td\xi} + \f{\td P}{\td\xi} =0,\quad \f{\p V}{\p y} = -\Xi_x\left(\f{\td U}{\td\xi} + \f{U}{\gamma P}\f{\td P}{\td\xi}\right),\quad \f{\td S}{\td\xi}=0,
\ee
where $R$ is the base flow density, given by $R=(P/\exp{S})^{1/\gamma}$. Substituting \eqref{decomp} into \eqref{eq:nonlinear} and retaining only the $\mathcal{O}(\varepsilon)$ terms results in the linearized disturbance equations,

\bse\label{eq:linearized}
\be\scalebox{1.0}{$
\f{\p u'}{\p \tau} + \Xi_x \f{\td U}{\td\xi} u' + U \Xi_x \f{\p u'}{\p \xi} 
+\f{\Xi_x}{R} \f{\p p'}{\p \xi} - \f{\Xi_x}{\gamma P R}\f{\td P}{\td\xi} p'
+ \f{\Xi_x }{\gamma R} \f{\td P}{\td\xi}\ s' +
\Xi_t \f{\td U}{\td\xi} \f{\td x_s'}{\td\tau} = 0,
$}
\ee
\be\scalebox{1.0}{$
\Xi_x \f{\td P}{\td\xi} u' 
+\gamma P \Xi_x \f{\p u'}{\p \xi} + 
\f{\p p'}{\p \tau} + U \Xi_x \f{\p p'}{\p \xi}
-\f{U\Xi_x}{P}\f{\td P}{\td\xi} p' + \f{\Xi_x^2}{2} \left(U\f{\td P}{\td\xi} + \gamma P \f{\td U}{\td\xi}\right)x_s' + \Xi_t \f{\td P}{\td\xi} \f{\td x_s'}{\td\tau} =0,
$}
\ee
\be\scalebox{1.0}{$
\f{\p s'}{\p \tau} + U \Xi_x \f{\p s'}{\p \xi} = 0,
$}
\ee
\ese
where the base flow vertical velocity derivative, $\p V/\p y$, is eliminated using \eqref{baseflow_eqs} and the disturbance vertical velocity derivative, $\p v'/\p y$, is assumed to be small and therefore neglected. The base flow metrics are given by

\be\scalebox{1.0}{$
\Xi_t = \f{\xi - 1 }{x_b-X_s},\quad \Xi_x = \f{2}{x_b-X_s}.
$}
\ee
The entropy equation is singular at the stagnation point, since $U(\xi=1)=0$ is a coefficient in front of $\p/\p\xi$. Therefore, a numerical solution is not straightforward. Nevertheless, a scheme dealing with the singularity is devised as detailed below to solve the equations accurately. First, the acoustic (isentropic) field is solved. Then, the entropic component is found by using an analytical expression for the singularity, which allows solving only for the entropy integration constant. A sketch of the problem, adapted from \citet{morkovin1960note}, is shown in Figure~\ref{sketch}(a).

\subsection{Entropy disturbance}

The entropy disturbances are found employing the following normal-mode ansatz,

\be\label{eq:s_modal}
s'(\xi, \tau) = \hat{s}(\xi) e^{-i\omega \tau} + \text{c.c.},
\ee
where $\hat{s}(\xi)$ represents the complex amplitude of the disturbance, $\omega$ is the frequency, and `c.c.' denotes the complex conjugate. The linearized entropy equation is

\be\label{eq:s_eq}
- i \omega \hat{s} + U\Xi_x\f{\td\hat{s}}{\td\xi} = 0.
\ee
The solution to this first-order linear differential equation is given by

\be\label{eq:s2_1}
\hat{s}(\xi) = \hat{s}_2 \hat{e}(\xi),\quad \hat{e}(\xi)=\exp\int_{-1}^{\xi} \f{i \omega}{\Xi_x U(\xi')} \, \td\xi',
\ee
where $\hat{s}_2$ is an integration constant determined by the linearized Rankine-Hugoniot shock jump conditions at $\xi=-1$, and $\hat{e}(\xi)$ is the normalized entropy mode shape, such that $\hat{e}(-1)=1$. Since the function is singular on the body ($\xi=1$), it is useful to rewrite it as

\be\label{eq:s2_s}
\hat{e}(\xi) = \left(\f{1-\xi}{2}\right)^n \exp\int_{-1}^{\xi} g(\xi') \, \td\xi', \quad g(\xi)=\f{n}{1-\xi}+\f{ i \omega }{\Xi_x U(\xi)},
\ee
where $n$ is chosen such that the integrand $g(\xi)$ becomes non-singular, and the integrand value on the body is found using L'Hôpital's rule, 

\be\label{eq:n}
n = \f{i\omega}{\Xi_x \f{\td U}{\td\xi}\big|_{\xi=1}},\quad \lim_{\xi \to 1}g(\xi)=-\f{i\omega}{2\Xi_x}\f{\f{\td^2U}{\td\xi^2}\big|_{\xi=1}}{\left(\f{\td U}{\td\xi}\big|_{\xi=1}\right)^2}.
\ee
Near the stagnation point $\hat{e}\sim(1-\xi)^n$, which attains values on the unit circle. The characteristic wavenumber of the oscillations, $\alpha \sim \omega / U$, tends to infinity as $U\to 0$. 

\begin{figure}[ht!]
\centering
\begin{subfigure}{0.45\linewidth}
\centering
\begin{overpic}[width=\linewidth]{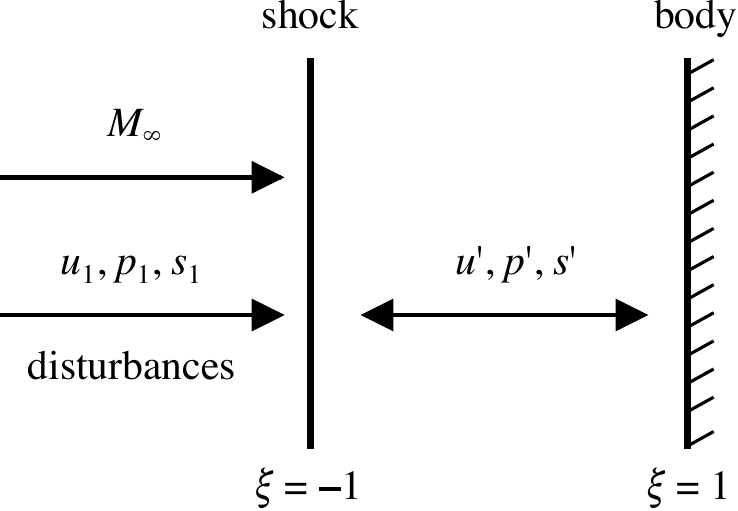}
\put(-5, 65){(a)}
\end{overpic}
\end{subfigure}
\hfill
\begin{subfigure}{0.45\linewidth}
\centering
\begin{overpic}[width=\linewidth]{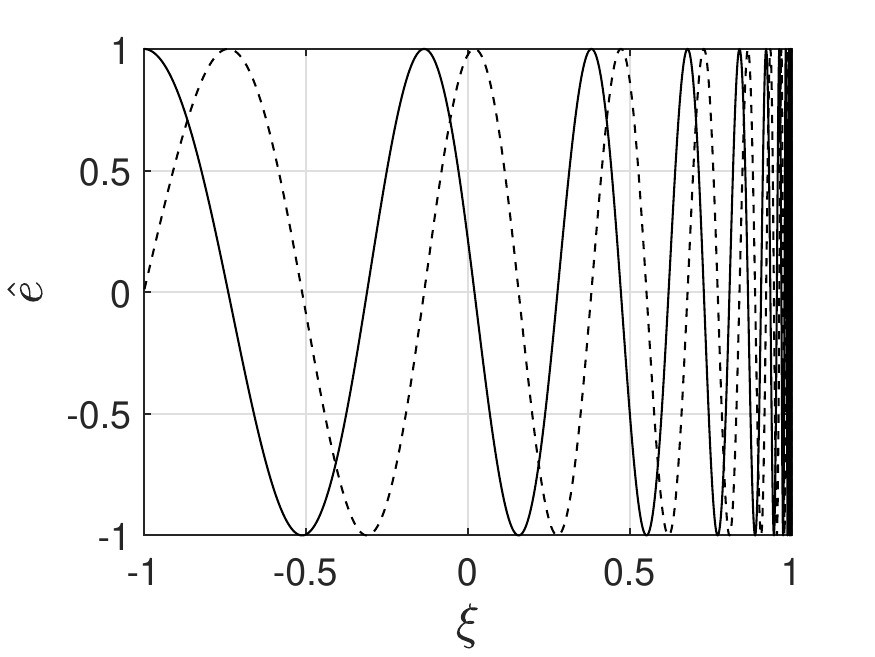}
\put(-5, 65){(b)}
\end{overpic}
\end{subfigure}
\caption{(a)~Sketch of the problem \citep[adapted from][]{morkovin1960note}. (b)~Schematic of the entropy wave $\hat{e}(\xi)$, solid/dashed lines indicate the real/imaginary parts, respectively.}
\label{sketch}
\end{figure}

A sample solution of the entropy equation is shown in Figure~\ref{sketch}(b). It can be seen that the wavenumber indeed increases towards the body (and the wavelength tends to zero). The only way to circumvent the singularity is by choosing $\hat{s}_2 = 0$, which means that there is no generation of entropy disturbances by the shock; however, this is nonphysical since the shock generates entropy disturbances. Note that inclusion of even a very small amount of viscosity is expected to regularize the singularity, since eventually the disturbance wavelength becomes small enough to be comparable to the viscous length scale.

\subsection{Acoustic disturbance}
Similarly to the entropy, a normal-mode ansatz is used,

\be\label{eq:modal}
\bq' = \tilde{\bq}(\xi) e^{-i\omega \tau} + (c.c.),
\ee
where $\omega$ is the frequency. Since, for the entropy, we only need to determine the jump across the shock $\hat{s}_2$, we define

\be\label{eq:qq}
\tilde{\bq} = (\hat{u}, \hat{p}, \hat{s}, \hat{x}_{s})^{\mathrm{T}} = \hat{\bq} \odot (1,\, 1,\, \hat{e},\, 1)^{\mathrm{T}}, \quad \hat{\bq} = (\hat{u}, \hat{p}, \hat{s}_2, \hat{x}_{s})^{\mathrm{T}},
\ee
which has two unknown functions, the velocity $\hat{u}(\xi)$ and pressure $\hat{p}(\xi)$, as well as two unknown scalars, the entropy jump $\hat{s}_2$ and the shock displacement $\hat{x}_s$. In addition, discretisation implies $\td/\td\xi=\Dxi$. The resulting equations are

\bse
\be\scalebox{0.9}{$
\left(-i\omega + \Xi_x \f{\td U}{\td\xi}+ U \Xi_x \Dxi \right) \hat{u} +
\left(\f{\Xi_x}{R} \Dxi - \f{\Xi_x}{\gamma P R}\f{\td P}{\td\xi} \right) \hat{p}
+ \left(\f{\Xi_x}{\gamma R} \f{\td P}{\td\xi} \hat{e} \right)\hat{s}_2 + \left(-i\omega \Xi_t \f{\td U}{\td\xi} \right) \hat{x}_s =0, 
$}
\ee
\be\scalebox{0.9}{$
\Xi_x\left(\f{\td P}{\td\xi} +\gamma P \Dxi\right) \hat{u}+
\left(-i\omega + U\Xi_x\left(\Dxi -\f{1}{P}\f{\td P}{\td\xi}\right)\right) \hat{p} 
+ \left(\f{\Xi_x^2}{2} \left(U\f{\td P}{\td\xi} + \gamma P \f{\td U}{\td\xi}\right) - i\omega \Xi_t \f{\td P}{\td\xi}\right)\hat{x}_s = 0.
$}
\ee
\ese
A special case arises when the base flow pressure gradient vanishes, i.e. $\td P/\td\xi = 0$. In this case, the equations governing the acoustic field become decoupled from the entropy disturbance, indicating that the post-shock acoustic (isentropic) and entropy (nonisentropic) modes evolve independently. This implies that the base flow pressure gradient acts as a coupling mechanism, scattering entropy disturbances into acoustic waves and vice versa.

\subsection{Combined solution}
The disturbance flow field is therefore written as a sum of two solutions,

\be\label{eq:decomp}
\hat{\bq} = \hat{\bq}_i + \hat{\bq}_{ii},
\ee
where the acoustic (isentropic) component, $\hat\bq_i=(\hat{u}_i, \hat{p}_i, 0, \hat{x}_{s,i})^\tT$, is obtained by enforcing $\hat s_{2,i}=0$, and the entropic component, $\hat{\bq}_{ii}=(\hat{u}_{ii}, \hat{p}_{ii}, \hat{s}_2, \hat{x}_{s,ii})^\tT$, satisfies $\hat s_{2,ii}=\hat s_2$. In the special case $\td P/\td\xi=0$, it follows that $\hat u_{ii}=\hat p_{ii}=\hat x_{s,ii}=0$, which means that the shock response is entirely driven by the acoustic mode, and the entropy perturbation is generated consequentially by the shock motion. 

\subsection{Boundary conditions}
The boundary conditions at the shock, derived from the linearized Rankine–Hugoniot relations, can be expressed in the following compact form. Upon applying the modal decomposition given in \eqref{eq:modal}, these conditions become

\be\label{eq:linear_RH}
\scalebox{0.9}{$
\begin{Bmatrix}
\hat{u}_2 \\[2pt]
\hat{p}_2 \\[2pt]
\hat{s}_2
\end{Bmatrix}
=
\bL
\begin{Bmatrix}
\hat{u}_1 \\[2pt]
\hat{p}_1 \\[2pt]
\hat{s}_1
\end{Bmatrix}
+ i\omega \bPi \hat{x}_s.
$}
\ee
Here, the subscript~2 refers to the state on the high pressure side of the shock, $\xi = -1$, while the subscript~1 denotes the incoming freestream disturbance on the low pressure side of the shock. The coefficient values for $\bL$ and $\bPi$ are provided in Appendix~\ref{AppB}. In our analysis, following the approach of \citet{morkovin1960note}, we consider only planar waves but examine the response to all three fundamental freestream wave types;

\bse
\be\scalebox{0.9}{$
\omega_e=\alpha \sqrt{\gamma} M_\infty,\quad\quad
\omega_f=\alpha \sqrt{\gamma}
\left(M_\infty+1\right),\quad
\omega_s=\alpha \sqrt{\gamma}
\left(M_\infty-1\right),
$}
\ee
\be\scalebox{0.9}{$
\begin{Bmatrix}
\hat{u}_1 \\[2pt]
\hat{p}_1 \\[2pt]
\hat{s}_1
\end{Bmatrix}_e =
\begin{Bmatrix}
0 \\[2pt]
0 \\[2pt]
-\gamma
\end{Bmatrix}
, \quad \quad 
\begin{Bmatrix}
\hat{u}_1 \\[2pt]
\hat{p}_1 \\[2pt]
\hat{s}_1
\end{Bmatrix}_f = 
\begin{Bmatrix}
1/\sqrt{\gamma} \\[2pt]
1 \\[2pt]
0 
\end{Bmatrix}
, \quad \quad 
\begin{Bmatrix}
\hat{u}_1 \\[2pt]
\hat{p}_1 \\[2pt]
\hat{s}_1
\end{Bmatrix}_s = 
\begin{Bmatrix}
-1/\sqrt{\gamma} \\[2pt]
1 \\[2pt]
0 
\end{Bmatrix},
$}
\ee
\be\scalebox{0.9}{$
\hat{\rho}_{1,e}=1,\quad\quad\quad\quad\quad\quad
\hat{\rho}_{1,f}=1/{\gamma},\quad\quad\quad\quad\quad
\hat{\rho}_{1,s}=1/{\gamma}.\quad\quad\quad
$}
\ee
\ese
Here, the subscripts $e$, $f$, and $s$ denote entropy, fast acoustic, and slow acoustic waves, respectively. The parameter $\alpha$ denotes the streamwise wavenumber and $\omega$ is the frequency. The density, $\hat{\rho}_1$, is obtained from the linearized entropy equation, $\hat{s}_1 = \hat{p}_1 - \gamma \hat{\rho}_1$. Any single-frequency planar disturbance wave can be expressed as a superposition of the three fundamental solutions above. The boundary conditions, following the decomposition in \eqref{eq:decomp}, are expressed in the following form,

\be\scalebox{0.89}{$
\begin{Bmatrix}
\hat{u}_{2,i} \\[2pt]
\hat{p}_{2,i} \\[2pt]
\hat{s}_{2,i}
\end{Bmatrix}
=
\begin{bmatrix}
\bL_{1\cdot} \\[2pt]
\bL_{2\cdot} \\[2pt]
\bs{0}
\end{bmatrix}
\begin{Bmatrix}
\hat{u}_1 \\[2pt]
\hat{p}_1 \\[2pt]
\hat{s}_1
\end{Bmatrix}
+ i\omega \begin{Bmatrix}
\Pi_{11} \\[2pt]
\Pi_{21} \\[2pt]
0
\end{Bmatrix} \hat{x}_{s,i},\quad
\begin{Bmatrix}
\hat{u}_{2,ii} \\[2pt]
\hat{p}_{2,ii} \\[2pt]
\hat{s}_{2,ii}
\end{Bmatrix}
=
\begin{bmatrix}
\bs{0} \\[2pt]
\bs{0} \\[2pt]
\bL_{3\cdot}
\end{bmatrix}
\begin{Bmatrix}
\hat{u}_1 \\[2pt]
\hat{p}_1 \\[2pt]
\hat{s}_1
\end{Bmatrix}
+ i\omega \begin{Bmatrix}
0 \\[2pt]
0 \\[2pt]
\Pi_{31}
\end{Bmatrix} \hat{x}_{s,i}
+ i\omega\bPi \hat{x}_{s,ii}.
$}
\ee
The remaining boundary condition is the no-penetration condition on the body,

\be
\hat{u}(\xi=1) = \hat{u}_i(\xi=1)= \hat{u}_{ii}(\xi=1) = 0.
\ee
The linearized equations were solved on a Chebyshev grid, with 250 collocation points.

\section{Results}
\begin{figure}[ht!]
\begin{center}
\begin{overpic}[width=\linewidth]{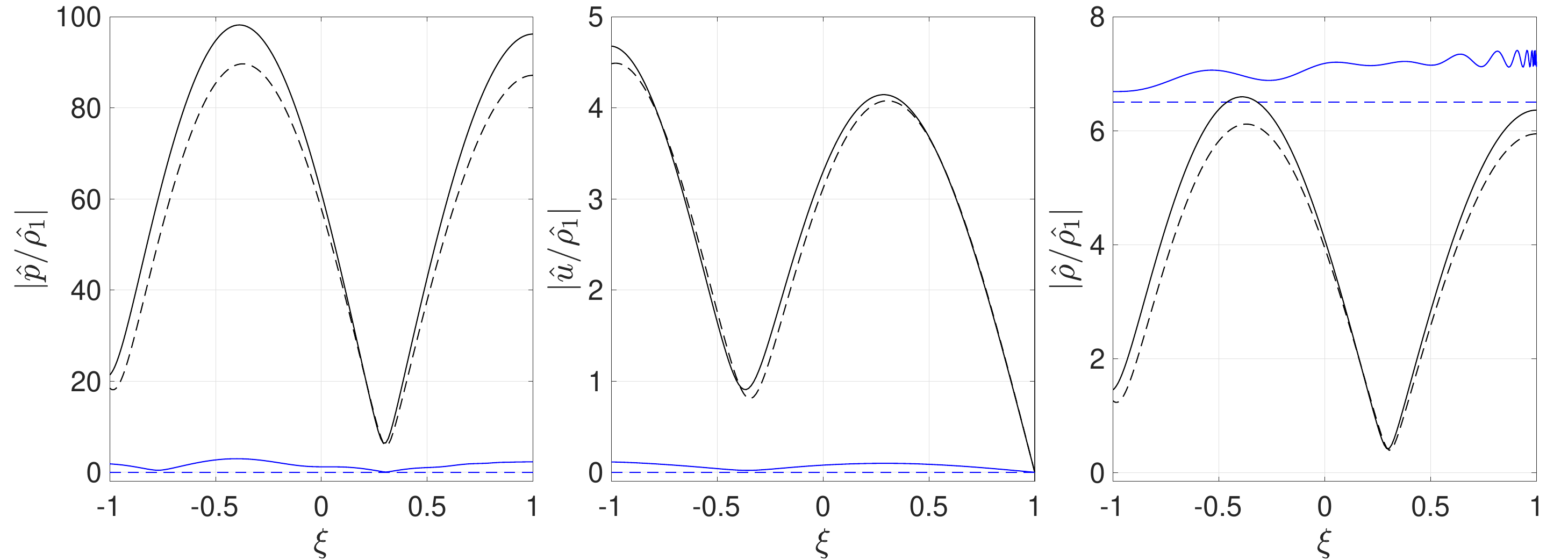}
\put(0, 35){(a)}
\put(34, 35){(b)}
\put(66, 35){(c)}
\end{overpic}
\caption{Acoustic $\square_i$ (black) and entropic $\square_{ii}$ ({\color{blue}{blue}}) components with (solid) and without (dashed) the base flow pressure gradient, for a cylinder at $M_\infty = 7$ and freestream entropy wave with $\omega = 40$ ($\tilde\omega = 2.9$). (a)~Pressure, (b)~Velocity, (c)~Density.}
\label{q_w_px}
\end{center}
\end{figure}

A cylinder at $M_\infty = 7$ is chosen as a representative case. Initially, the response of the flow on the stagnation line to an incoming entropy wave with frequency $\omega = 40$ is examined. Then, a frequency sweep is conducted, followed by a variation of the Mach number. The amplitudes of $\hat{p}$, $\hat{u}$ and $\hat{\rho}$ for the acoustic ($\hat{\bq}_i$) and entropic ($\hat{\bq}_{ii}$) components are shown by the solid black and blue lines in Figure~\ref{q_w_px}, respectively, with the density perturbation, $\hat{\rho}$, obtained from $\hat{s} = \hat{p}/P - \gamma \hat{\rho}/R$. Although the total density perturbation includes comparable acoustic and entropic contributions, the pressure and velocity disturbances are dominated by the acoustic (isentropic) mode, for which $\hat{\rho} = (R / \gamma P)\hat{p}$. Therefore, it is justified to disregard the entropic component when the velocity and pressure fields are of primary interest.

The base flow pressure gradient, $\td P/\td\xi$, introduces a coupling between the acoustic and entropic modes, which is quantified by comparison with the solution assuming $\td P/\td\xi=0$, indicated by the dashed lines. The solution neglecting $\td P/\td\xi$ assumes uniform pressure and density while retaining the actual velocity distribution. Only minor quantitative changes in the solution are observed. Therefore, omitting the term $\td P/\td\xi$ provides a justified simplification that preserves the essential characteristics of the flow while reducing the analytical complexity. The shock displacement amplitudes are $|\hat{x}_{s,i}/\hat{\rho}_1| = 0.142$ and $|\hat{x}_{s,ii}/\hat{\rho}_1| = 0.003$, indicating that the acoustic component dominates the shock response. In the absence of the base flow pressure gradient, only the acoustic response remains, with $|\hat{x}_s/\hat{\rho}_1|=|\hat{x}_{s,i}/\hat{\rho}_1| = 0.136$.

The frequency shown in figure~\ref{q_w_px} corresponds to the second peak that maximizes the disturbance pressure on the body (see figure~\ref{pb_and_f_vs_omega} and related discussion in the next subsection). For this frequency, there is a constructive interference between the fast and slow acoustic waves, that can be quantified by the ratio between the values of the acoustic pressure and density on the body and the shock, yielding $\hat{p}_{b,i}/\hat{p}_{2,i} \approx \hat{\rho}_{b,i}/\hat{\rho}_{2,i} \approx 4.5$. The entropic density component, however, grows only slightly from the shock to the body due to the rise in base flow density, and the oscillatory behavior of the entropy wave (figure~\ref{sketch}(b)) is visible as $\xi\to1$. Since the velocity perturbations of the acoustic waves are of opposite phases, the acoustic velocity disturbance attains local minima where the acoustic pressure attains local maxima and vice versa. For the frequency of the first peak that maximizes the disturbance pressure on the body, monotonic increase (decrease) of the acoustic pressure (velocity) is observed, with the extrema attained at the boundaries (not shown here).

\subsection{Effect of frequency}
\begin{figure}[ht!]
\centering
\begin{subfigure}{0.49\linewidth}
\centering
\begin{overpic}[width=\linewidth]{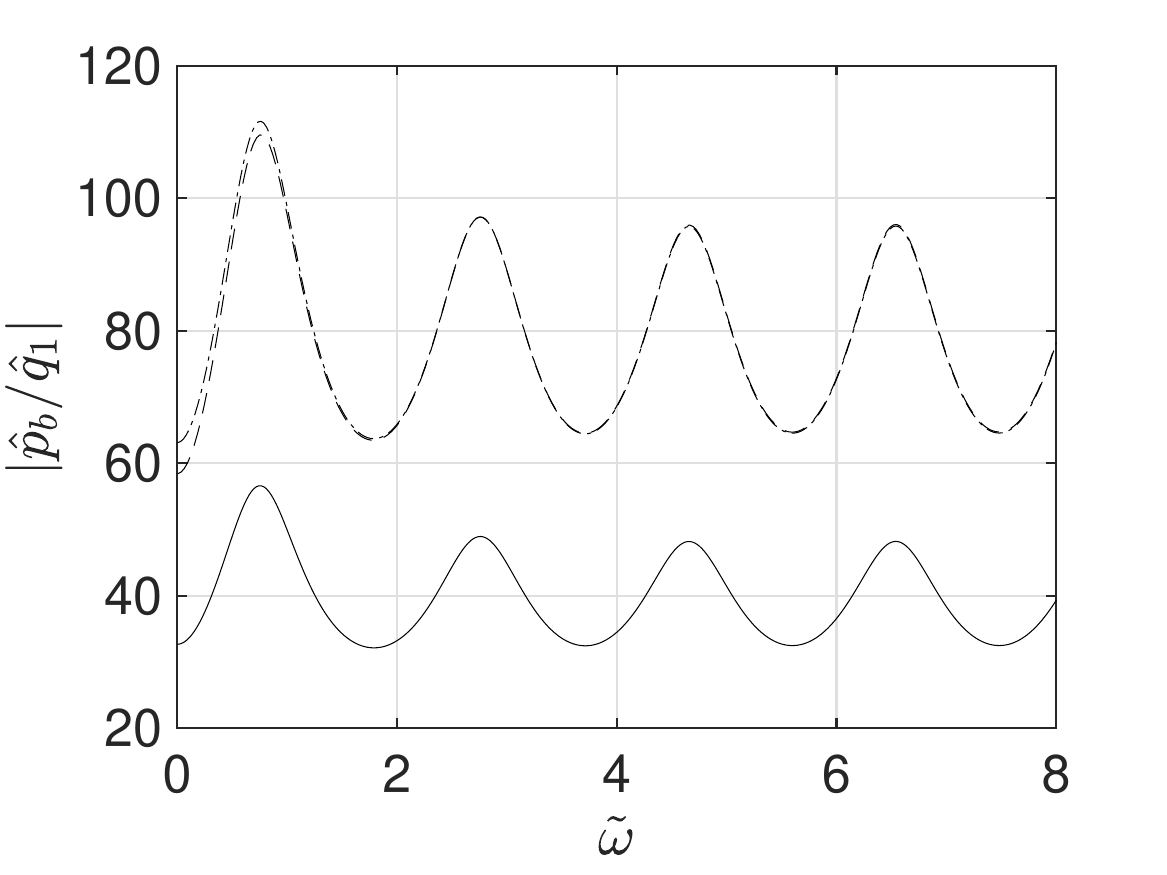}
\put(-1, 70){(a)}
\end{overpic}
\end{subfigure}
\hfill
\begin{subfigure}{0.49\linewidth}
\centering
\begin{overpic}[width=\linewidth]{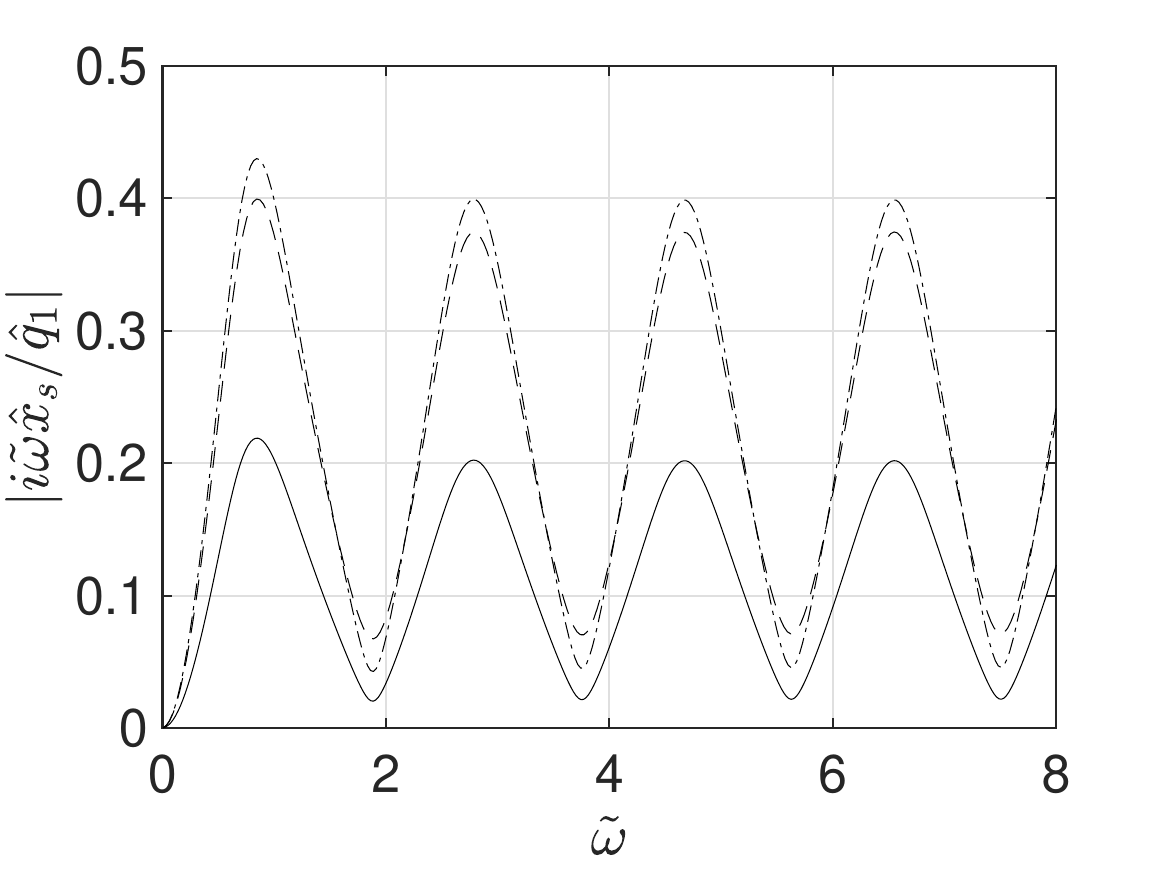}
\put(-1,70){(b)}
\end{overpic}
\end{subfigure}
\caption{(a)~Disturbance pressure amplitude on the body, $|\hat{p}_b|$, and (b)~Shock velocity amplitude, $|i\tilde{\omega}\hat{x}_s|$, vs. rescaled frequency $\tilde{\omega}$, for incoming slow acoustic (solid); entropy (dashed-dotted); and fast acoustic (dashed) waves, for a cylinder at $M_\infty = 7$. }
\label{pb_and_f_vs_omega}
\end{figure}

The effect of frequency is assessed by examining the pressure disturbance amplitude on the body, $|\hat{p}_b|$, and the shock velocity amplitude, $|i\tilde{\omega}\hat{x}_s|$, shown in Figures \ref{pb_and_f_vs_omega}(a) and (b), respectively. The values are normalized by the freestream disturbance amplitude $|\hat{q}_1|$, which for the acoustic modes is $\hat{q}_1 = \hat{p}_1$, while for the entropy mode it is $\hat{q}_1 = \hat{\rho}_1$. The frequency is rescaled as follows

\be\label{eq:tilde_omega}
\tilde{\omega} = \frac{2 \, \omega \, \Delta}{\pi \sqrt{\gamma P_0 / R_0}},
\ee 
where $\Delta$ is the shock standoff distance, $P_0$ is the stagnation pressure, and $R_0$ is the stagnation density. This renormalization, as obtained from the analytical solution presented in \citet{morkovin1960note}, places the extrema of the disturbance pressure on the body at integer values of $\tilde{\omega}$ for uniform post-shock flows. In the present results for $M_\infty = 7$, a shift in the location of these extrema is observed compared to the prediction based on uniform flow. The first peak, which occurs at $\tilde\omega\approx0.75$, in both pressure and shock velocity amplitudes, is the highest, with comparable amplification of fast acoustic disturbances (dashed lines) and  entropy waves (dashed-dotted lines). Slow acoustic waves (solid lines) reach about 50\% of the response to fast acoustic waves.

\subsection{Effect of Mach number}
\begin{figure}[ht!]
\begin{center}
\includegraphics[width=0.7\linewidth]{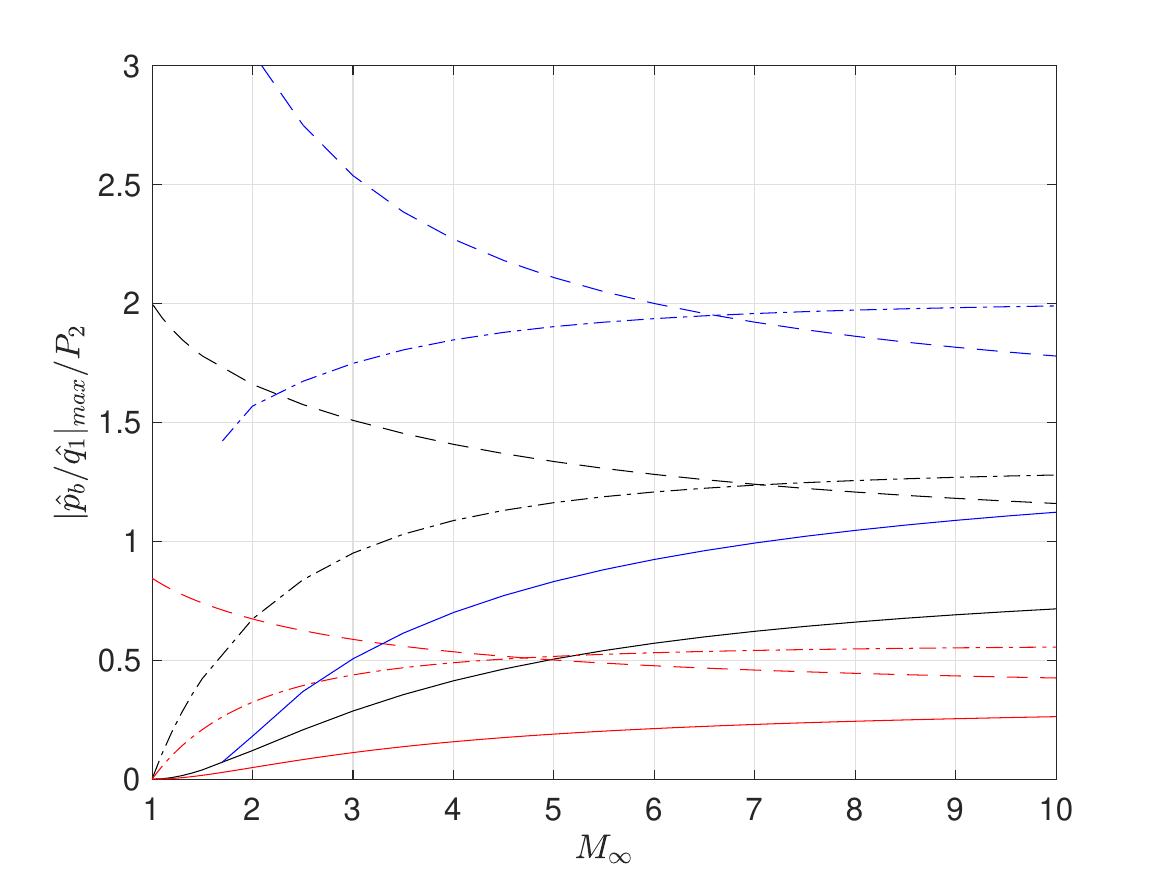}
\caption{Maximum amplitude of the disturbance pressure on the body, $|\hat{p}_b|_{max}$, vs. Mach number. Actual flow ({\color{blue}blue}), uniform flow (black) and no reflections ({\color{red}red}). Incoming slow acoustic (solid); entropy (dashed-dotted); and fast acoustic (dashed) waves.}
\label{pb_max_all}
\end{center}
\end{figure}

The effect of the freestream Mach number, $M_\infty$, is quantified by examining the maximum amplitude of the disturbance pressure on the body, $|\hat{p}_b|_{max}$, attained at $\tilde\omega\approx0.75$. The value of $|\hat{p}_b/\hat{q}_1|_{max}$, normalized by the base flow post-shock pressure $P_2$, is shown in Figure~\ref{pb_max_all} for the three types of freestream disturbances: slow acoustic (solid lines), entropy (dashed-dotted lines), and fast acoustic (dashed lines). Three different configurations are compared: actual decelerating stagnation line flows are colored blue, the simplified uniform post-shock flow proposed by \citet{morkovin1960note} is colored black, and the response of the shock neglecting reflections from the body as analyzed by \citet{Ribner54a} is colored red. The results show that, as expected, the presence of the body always leads to amplification of the incoming disturbances. The actual flow (blue) leads to about 50\% higher amplification compared to the uniform flow (black). Among the disturbance types, for Mach numbers smaller than approximately 6.6 the fast acoustic wave is the most highly amplified, followed by the entropy wave and finally the slow acoustic wave. For Mach numbers above approximately 6.6 the entropy wave surpasses the fast acoustic wave. The amplification trends with the Mach number vary between the slow acoustic and entropy waves, which increase with $M_\infty$, and the fast acoustic wave, which decreases with $M_\infty$. All responses level off at higher Mach numbers, as $M_\infty\to\infty$.

\section{Discussion and Conclusions}
The receptivity problem targeted at the flow on the stagnation line of blunt bodies in supersonic flow is formulated and solved by combining a spectral shock-fitting algorithm with a linear disturbance analysis. While reproducing the results of \citet{morkovin1960note} for uniform post-shock flow, a wrong sign was identified in one of the coefficients, inadvertently suggesting damping by the body; with the corrected sign, the presence of the body amplifies disturbances. Our main findings are as follows,
\begin{itemize}
\item Acoustic disturbances dominate pressure and velocity responses, while the density response is influenced by both acoustic and entropy modes.
\item Post-shock entropy disturbances exhibit singular behavior near the stagnation point, with their wavenumber stretching to infinity.
\item The base flow pressure gradient introduces a coupling between acoustic and entropic disturbances; however, since these are relatively minor corrections, it is justified to neglect the base flow pressure gradient.
\item The maximal disturbance pressure on the body occurs when the rescaled frequency $\tilde\omega$ is approximately 0.75.
\item The realistic stagnation line base flow amplifies all disturbance types about 50\% higher compared to the idealized uniform post-shock flow.
\item The amplification of fast acoustic waves decreases with $M_\infty$, while the response to entropy and slow acoustic waves increases with $M_\infty$. The most amplified disturbance changes from the fast acoustic to the entropy wave at $M_\infty\approx6.6$. All responses level off as the Mach number increases.
\end{itemize}

The applicability of the one-dimensional model to actual two-dimensional (planar or axisymmetric) flows is an important subject for future research. In addition, the interaction of the shock with vorticity disturbances, which do not exist in one-dimensional flow, and disturbances which are oblique with respect to the stagnation streamline, also require further investigation.


\section*{Funding}
The research was supported by an internal grant from Technion Israel Institute of Technology.

\section*{Declaration of Interests}
The authors report no conflict of interest.

\section*{Author ORCIDs}
Iliya Milman http://orcid.org/0009-0006-8938-2253 \\
Michael Karp http://orcid.org/0000-0003-3649-2492

\clearpage

\appendix

\section{Discussion and correction of the analysis by Morkovin}\label{AppA}

It should be noted that all the mathematical expressions in \citet{morkovin1960note} are correct. However, the plots and conclusions drawn from them are not, due to an inadvertently added minus sign to~$\Pi_{31}$. This coefficient,

\be\label{Pi31}
\Pi_{31} = \f{1}{1 - M^2}\left(1 - \f{\rho_{1m}}{\rho_{m}}\right)
\left(1 + M^2 + (\gamma - 1)\left(1 - \f{\rho_{m}}{\rho_{1m}}\right) M^2\right),
\ee
is always positive; however, throughout \citet{morkovin1960note} it was consistently used with a minus sign. In the above expression $M$ is the post-shock Mach number and $\rho_{m}/\rho_{1m}$ is the density ratio across the shock. Note that all the expressions in this appendix refer to their definitions in \citet{morkovin1960note}.

\begin{figure}[ht!]
\begin{center}
\includegraphics[width=0.7\linewidth]{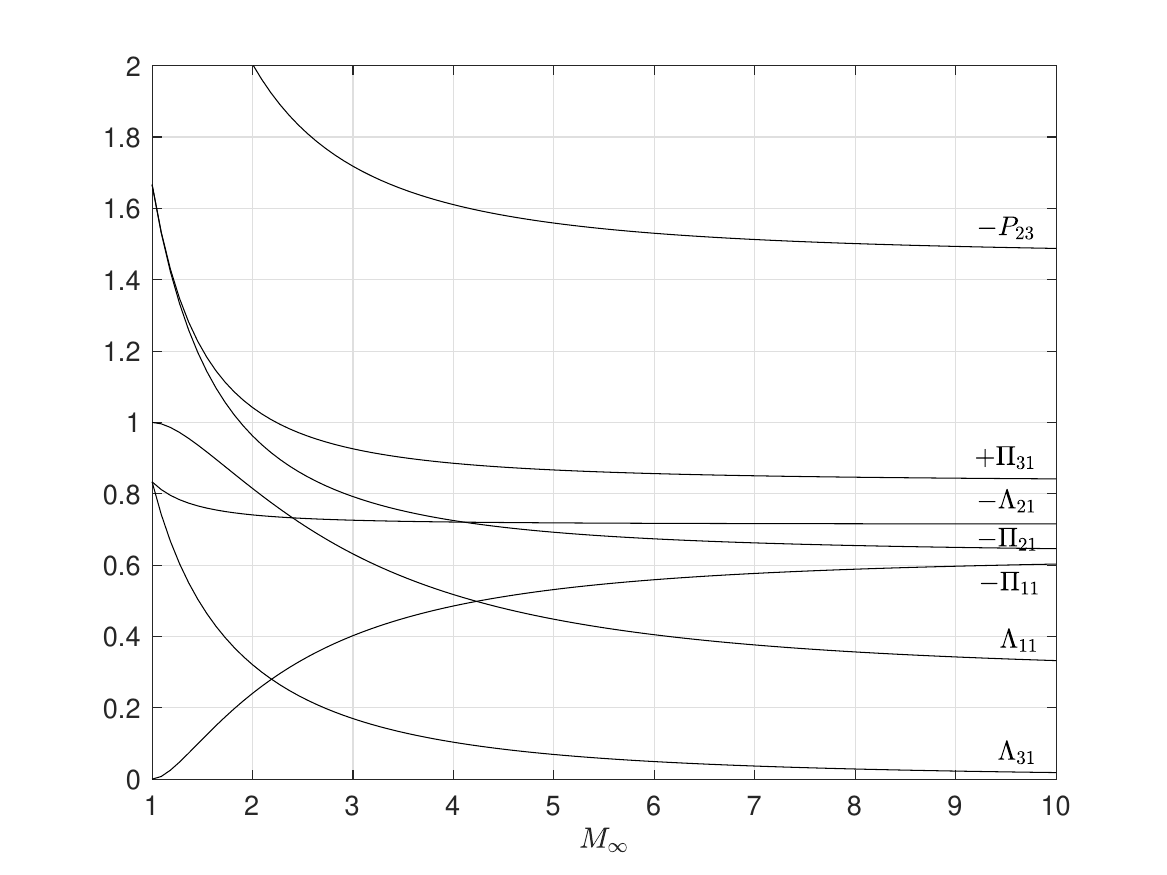}
\caption{Corrected figure~2 from \citet{morkovin1960note}, with $+\Pi_{31}$ and $-P_{23}$.}
\label{mork1}
\end{center}
\end{figure}

\begin{figure}[ht!]
\begin{center}
\includegraphics[width=0.7\linewidth]{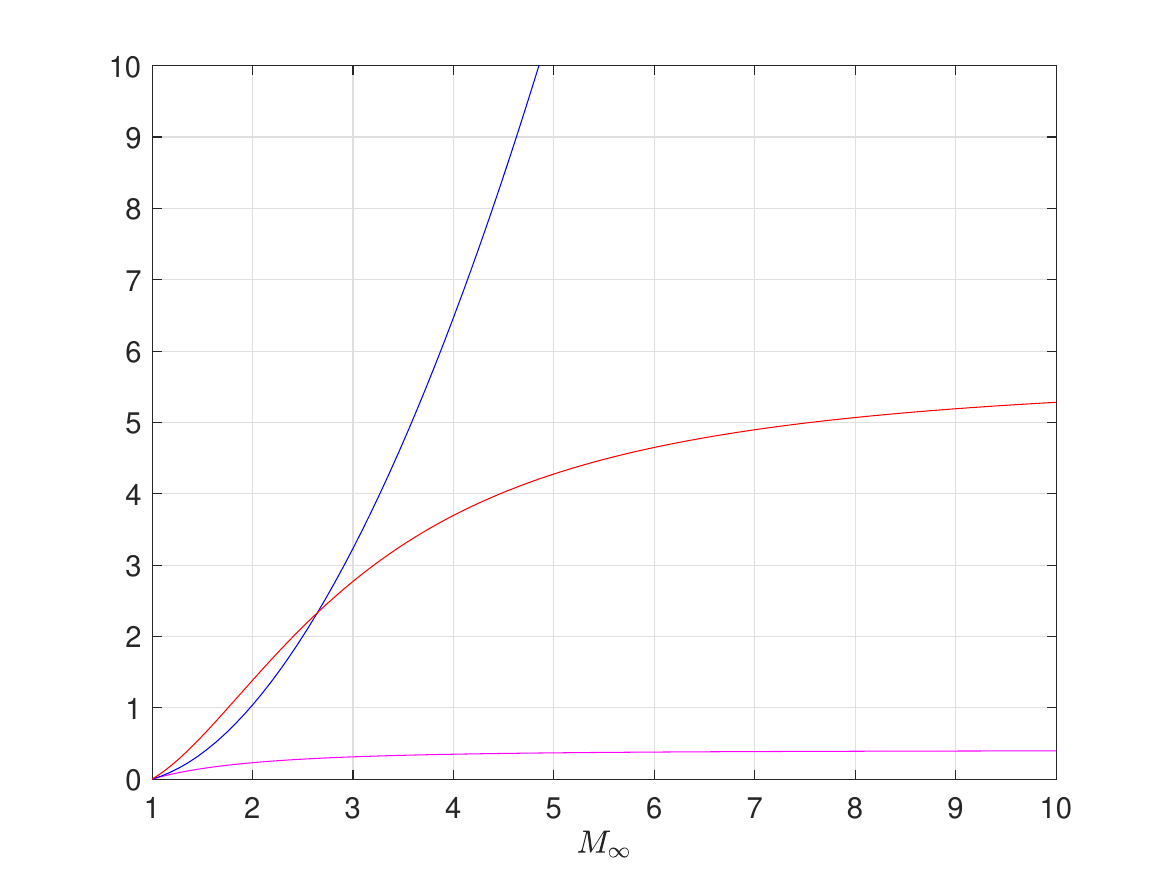}
\caption{Corrected figure~3 from \citet{morkovin1960note}, (8a)~({\color{blue}blue}); $\biggl|\displaystyle \frac{\delta p_+}{\f{\gamma}{2} p_m M^2}\Big/\frac{\delta T_-}{T_{1m}}\biggr|$ ({\color{red}red}); $|\f{\Delta_{23}}{P_{23}}|$~(7a) ({\color{magenta}magenta}).}
\label{mork2}
\end{center}
\end{figure}

\begin{figure}[ht!]
\begin{center}
\includegraphics[width=0.7\linewidth]{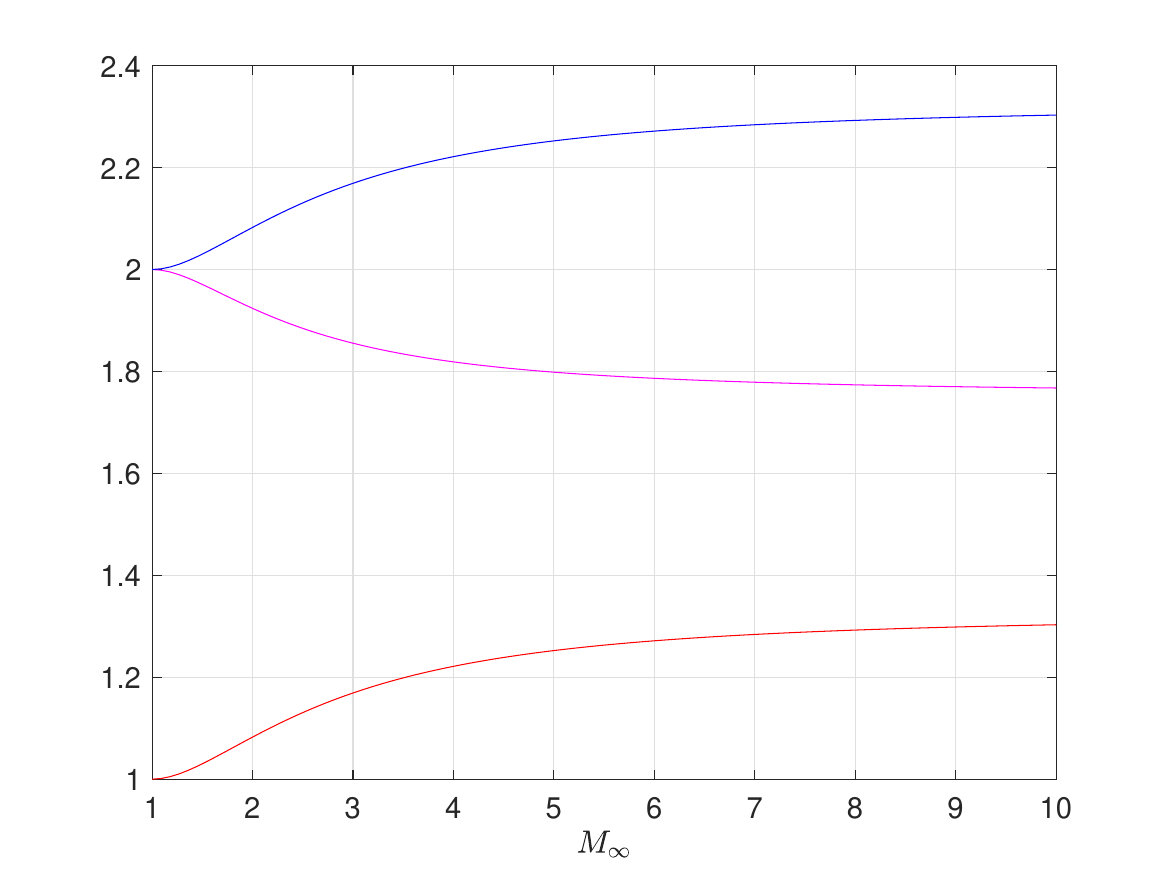}
\caption{Corrected figure~5 from \citet{morkovin1960note}, $|\frac{P_{23}}{\Pi_{21}}|$~(13b) ({\color{blue}blue}); $|\frac{P_{23}}{\Pi_{31}}|$~(13c) ({\color{magenta}magenta}); max of~(14) $|\frac{\Pi_{31}}{\Pi_{21}}|$ ({\color{red}red}).}
\label{mork3}
\end{center}
\end{figure}

The corrected versions of figures 2, 3 and 5 in \citet{morkovin1960note} (without the inadvertently added minus sign to $\Pi_{31}$) are shown in figures~\ref{mork1}, \ref{mork2}, and \ref{mork3}, respectively. In figure~\ref{mork1} only the curves of $\Pi_{31}$ and $P_{23}$ are different from the original \citep[figure~2 in][]{morkovin1960note}. Note that $P_{23}=\Pi_{21}-\Pi_{31}$ is negative, therefore, a minus sign is added to it to consistently plot only positive quantities. In figure~\ref{mork2} all coefficients smoothly decrease to 0 as $M_\infty\to1$, compared to the divergent behavior in the original \citep[figure~3 in][]{morkovin1960note}. In figure~\ref{mork3}, the values of $|{P_{23}}/{\Pi_{21}}|$ and $|{P_{23}}/{\Pi_{31}}|$, which are the multiplying factors as a result of the body, are close to~2, which implies that reflections from the body lead to amplification of disturbances, compared to values lower than unity in the original \citep[figure~5 in][]{morkovin1960note}.

\section{Coefficient expressions}\label{AppB}

The expressions for the coefficients in \eqref{eq:linear_RH} are given in Table~\ref{tab:coef}.

\begin{table}
\centering
\caption{Coefficients $\Lambda_{ij}$ and $\Pi_{ij}$.}
\label{tab:coef}
\begin{tabular}{c c | c c}
\textbf{Coefficient} & \textbf{Expression} & \textbf{Coefficient} & \textbf{Expression} \\[8pt]
$\Lambda_{11}$ & $-\dfrac{(\gamma - 1) M_{\infty}^2 - 2}{(\gamma + 1) M_{\infty}^2}$ &
$\Lambda_{31}$ & $\dfrac{4 (\gamma - 1) \sqrt{\gamma} (M_{\infty}^2 - 1)^2}{M_{\infty} \left( 2 + (\gamma - 1) M_{\infty}^2 \right) \left( 1 - \gamma + 2 \gamma M_{\infty}^2 \right)}$ \\[8pt]
$\Lambda_{12}$ & $-\dfrac{2 (\gamma - 1)}{M_{\infty} (\gamma + 1) \sqrt{\gamma}}$ &
$\Lambda_{32}$ & $-\dfrac{2 (\gamma - 1)^2 (M_{\infty}^2 - 1)^2}{\left( 2 + (\gamma - 1) M_{\infty}^2 \right) \left( 1 - \gamma + 2 \gamma M_{\infty}^2 \right)}$ \\[8pt]
$\Lambda_{13}$ & $-\dfrac{2}{M_{\infty} (\gamma + 1) \sqrt{\gamma}}$ &
$\Lambda_{33}$ & $ 1 - \dfrac{2(\gamma - 1)(M_{\infty}^2 - 1)^2}{(2 + (\gamma - 1) M_{\infty}^2)(1 - \gamma + 2 \gamma M_{\infty}^2)} $\\[8pt]
$\Lambda_{21}$ & $\dfrac{4 \sqrt{\gamma} M_{\infty}}{\gamma + 1}$ &
$\Pi_{11}$ & $-\dfrac{2(M_{\infty}^2 + 1)}{M_{\infty}^2 (\gamma + 1)}$ \\[8pt]
$\Lambda_{22}$ & $\dfrac{2 M_{\infty}^2 - \gamma + 1}{\gamma + 1}$ &
$\Pi_{21}$ & $-\dfrac{4 \sqrt{\gamma} M_{\infty}}{\gamma + 1}$ \\[8pt]
$\Lambda_{23}$ & $-\dfrac{2 M_{\infty}^2}{\gamma + 1}$ &
$\Pi_{31}$ & $-\dfrac{4 (\gamma - 1) \sqrt{\gamma} (M_{\infty}^2 - 1)^2}{M_{\infty} \left( 2 + (\gamma - 1) M_{\infty}^2 \right) \left( 1 - \gamma + 2 \gamma M_{\infty}^2 \right)}$ \\[8pt]
\end{tabular}
\end{table}

\bibliography{bib}

@article{morkovin1960note,
    author = {Morkovin, M. V.},
    title = {Note on the Assessment of Flow Disturbances at a Blunt Body Traveling at Supersonic Speeds Owing to Flow Disturbances in Free Stream},
    journal = {J. Appl. Mech.},
    volume = {27},
    number = {2},
    pages = {223-229},
    year = {1960},
    month = {06},
    issn = {0021-8936},
    doi = {10.1115/1.3643942},
    url = {https://doi.org/10.1115/1.3643942}
}

@article{brooks2004,
  author = {Brooks, G. P. and Powers, J. M.},
  title = {Standardized Pseudospectral Formulation of the Inviscid Supersonic Blunt Body Problem},
  journal = {J. Comput. Phys.},
  volume = {197},
  number = {1},
  year = {2004},
  month = {June},
  pages = {58--85},
  doi = {10.1016/j.jcp.2003.11.017},
  url = {https://doi.org/10.1016/j.jcp.2003.11.017}
}

@article{kopriva1999,
  author = {Kopriva, D. A.},
  title = {Shock-Fitted Multidomain Solution of Supersonic Flows},
  journal = {Comput. Methods in Appl. Mech. Eng.},
  volume = {175},
  number = {3--4},
  year = {1999},
  month = {July},
  pages = {383--394},
  doi = {10.1016/S0045-7825(98)00362-4},
}

@article{najafi2014shockfit,
  author  = {Najafi, M. and Hejranfar, K. and Esfahanian, V.},
  title   = {Application of a shock-fitted spectral collocation method for computing transient high-speed inviscid flows over a blunt nose},
  journal = {J. Comput. Phys.},
  volume  = {257},
  year    = {2014},
  pages   = {954--980},
  doi     = {10.1016/j.jcp.2013.09.037},
}

@article{Ribner54a,
title={Convection of a pattern of vorticity through a shock wave},
author={Ribner, H. S.},
journal = {NACA-TR-1164},
year = {1954},
pages={1--17},
url={https://ntrs.nasa.gov/api/citations/19930092192/downloads/19930092192.pdf}
}

@article{Ribner54c,
title={Shock-turbulence interaction and the generation of noise},
author={Ribner, H. S.},
journal = {NACA-TR-1233},
year={1954},
pages={683--701},
url={https://ntrs.nasa.gov/api/citations/19930090990/downloads/19930090990.pdf}
}

@article{McKenzieWestphal,
author = {McKenzie, J. F. and Westphal, K. O.},
title = {Interaction of Linear Waves with Oblique Shock Waves},
journal = {Phys. Fluids},
volume = {11},
pages = {2350--2362},
year = {1968},
doi = {10.1063/1.1691825}
}

@article{Moore54,
author={Moore, F. K.},
title={Unsteady oblique interaction of a shock wave with a plane disturbance},
journal = {NACA-TR-1165},
year = {1954},
pages={1--21},
url={https://ntrs.nasa.gov/api/citations/19930092193/downloads/19930092193.pdf}
}

@article{MilmanKarp,
	title = {Mach {Independence} of {Entropy} {Layer} {Instabilities}},
	volume = {39},
	issn = {1432-2250},
	url = {https://doi.org/10.1007/s00162-025-00758-w},
	doi = {10.1007/s00162-025-00758-w},
	number = {39},
	journal = {Theor. Comput. Fluid Dyn.},
	author = {Milman, I. and Karp, M.},
	year = {2025},
	pages = {1--16},
}

@article{Xiong2023,
  author       = {Xiong, Y. and Zhao, L. and Wu, J.},
  title        = {Characterization of Disturbance Resonance in Postshock of Blunt Body in Hypersonic Flow},
  journal      = {AIAA J.},
  year         = {2023},
  volume={61},
  number={9},
  pages={3735--3742},
  doi          = {10.2514/1.J062615},
}

@article{Chaudhry2017,
  author        = {Chaudhry, R. S. and Candler, G. V.},
  title         = {Computing measured spectra from hypersonic pitot probes with flow-parallel freestream disturbances},
  journal       = {AIAA J.},
  volume        = {55},
  number        = {12},
  pages         = {4155--4166},
  year          = {2017},
  doi           = {10.2514/1.J055396}
}

@article{SchildenJFM17,
title={Numerical analysis of high speed wind tunnel flow disturbance measurements using stagnation point probes}, 
author={Schilden, T. and Schröder, W.},
journal={J. Fluid Mech.},
year={2017},
volume={833}, 
pages={247–273},
DOI={10.1017/jfm.2017.674}
}

@article{SchildenJFM19,
title={Inclined slow acoustic waves incident to stagnation point probes in supersonic flow},
author={Schilden, T. and Schröder, W.},
journal={J. Fluid Mech.},
year={2019},
volume={866},
pages={567–597},
DOI={10.1017/jfm.2019.121}
}

@article{sawant22,
  title={On the synchronisation of three-dimensional shock layer and laminar separation bubble instabilities in hypersonic flow over a double wedge},
  author={Sawant, S. S. and Theofilis, V. and Levin, D. A.},
  journal={J. Fluid Mech.},
  volume={941},
  pages={A7},
  year={2022},
  publisher={Cambridge University Press},
  DOI = {10.1017/jfm.2022.276}
}

\end{document}